\newcommand{\sx}{\sigma^x}
\newcommand{\sy}{\sigma^y}
\newcommand{\sz}{\sigma^z}

\newcommand{\U}{\uparrow}
\newcommand{\D}{\downarrow}

\documentclass[%
reprint,amsmath,amssymb,aps,prx,superscriptaddress,longbibliography,
]{revtex4-1}

\usepackage{graphicx}
\usepackage{dcolumn}
\usepackage{bm}
\usepackage{epstopdf}
\usepackage{float}
\usepackage{braket}
\usepackage{dsfont}
\usepackage{amsmath}
\usepackage{xcolor}

\usepackage{lipsum}

\begin{document}

\title{Coherent multi-spin exchange coupling in a quantum-dot spin chain}

\author{Haifeng Qiao}
\thanks{These authors contributed equally.}

\author{Yadav P. Kandel}
\thanks{These authors contributed equally.}

\affiliation{Department of Physics and Astronomy, University of Rochester, Rochester, NY, 14627 USA}

\author{Kuangyin Deng}
\affiliation{Department of Physics, Virginia Tech, Blacksburg, Virginia 24061, USA}

\author{Saeed Fallahi}
\affiliation{Department of Physics and Astronomy, Purdue University, West Lafayette, IN, 47907 USA}
\affiliation{Birck Nanotechnology Center, Purdue University, West Lafayette, IN, 47907 USA}

\author{Geoffrey C. Gardner}
\affiliation{Birck Nanotechnology Center, Purdue University, West Lafayette, IN, 47907 USA}
\affiliation{School of Materials Engineering, Purdue University, West Lafayette, IN, 47907 USA}

\author{Michael J. Manfra}
\affiliation{Department of Physics and Astronomy, Purdue University, West Lafayette, IN, 47907 USA}
\affiliation{Birck Nanotechnology Center, Purdue University, West Lafayette, IN, 47907 USA}
\affiliation{School of Materials Engineering, Purdue University, West Lafayette, IN, 47907 USA}
\affiliation{School of Electrical and Computer Engineering, Purdue University, West Lafayette, IN, 47907 USA}

\author{Edwin Barnes}
\affiliation{Department of Physics, Virginia Tech, Blacksburg, Virginia 24061, USA}

\author{John M. Nichol}
\email{john.nichol@rochester.edu}

\affiliation{Department of Physics and Astronomy, University of Rochester, Rochester, NY, 14627 USA}

\begin{abstract}
Heisenberg exchange coupling between neighboring electron spins in semiconductor quantum dots provides a powerful tool for quantum information processing and simulation. Although so far unrealized, extended Heisenberg spin chains can enable long-distance quantum information transfer and the generation of non-equilibrium quantum states. In this work, we implement simultaneous, coherent exchange coupling between all nearest-neighbor pairs of spins in a quadruple quantum dot. The main challenge in implementing simultaneous exchange couplings is the nonlinear and nonlocal dependence of the exchange couplings on gate voltages. Through a combination of electrostatic simulation and theoretical modeling, we show that this challenge arises primarily due to lateral shifts of the quantum dots during gate pulses. Building on this insight, we develop two models, which can be used to predict the confinement gate voltages for a desired set of exchange couplings. Although the model parameters depend on the number of exchange couplings desired (suggesting that effects in addition to lateral wavefunction shifts are important), the models are sufficient to enable simultaneous and independent control of all three exchange couplings in a quadruple quantum dot. We demonstrate two-, three-, and four-spin exchange oscillations, and our data agree with simulations. 
\end{abstract}


\pacs{}

\maketitle
\section{Introduction}
A unique and valuable feature of electron-spin qubits in quantum dots~\cite{Loss1998,Kane1998} is the voltage-controlled nearest-neighbor Heisenberg exchange coupling. Heisenberg exchange coupling results from the interplay of the electronic confinement potential, the Coulomb interaction, and the antisymmetric nature of the electronic wavefunctions under particle exchange. On a basic level, exchange coupling enables two-~\cite{Loss1998,DiVincenzo2000ExchangeQC,Nowack2011SwapGate,Zajac2017CNot} and three-qubit gates~\cite{Gullans2019Toffoli} for single-spin qubits. Exchange coupling also allows rapid and high-fidelity initialization and readout of pairs of spins. As a result, exchange coupling underlies the operation of electron spin qubits consisting of two~\cite{Petta2005STQubit,Foletti2009}, three ~\cite{Laird2010ExchangeOnly,Medford2013ExchangeOnly,Shi2012,Kim2014,Eng2014,Shim2016}, or more~\cite{Sala2017,Russ2018ExchangeOnly,Sala2019} electrons. Superexchange~\cite{Baart2016S,Malinowski2019} in multi-electron systems and extended exchange-coupled spin chains can enable new forms of quantum-information transfer ~\cite{Bose2003SpinChain,Bose2007} and the generation of many-body entangled states~\cite{Friesen2007SpinBus}. Recent experiments exploiting pulsed exchange coupling in spin chains point to the feasibility of these proposals~\cite{Kandel2019SpinTransfer}. Heisenberg spin chains are also predicted to generate non-equilibrium quantum phenomena~\cite{Barnes2016MBL,Barnes2019TimeCrystal}. 

In part due to these exciting possibilities, independent and automated control of inter-dot tunnel couplings has been the focus of intense research in quantum-dot arrays \cite{Hensgens2017,Mukhopadhyay2018,Sigillito2019,vanDiepen2018,Mills2019Computer,Hsiao2020}. However, generating multiple independent, non-zero, and coherent exchange couplings in quantum-dot arrays is challenging for several reasons. First, the standard procedure to measure tunnel couplings involves detuning pairs of dots away from the symmetric idling point~\cite{DiCarlo2004}, making it difficult to calibrate exchange couplings under actual experimental conditions. Second, as discussed further below, the nonlinear and nonlocal dependence of the exchange couplings on the confinement gate voltages poses a significant challenge. Third, multiple non-zero exchange couplings generate complicated spectra that do not permit easy measurement and iterative tuning of individual exchange frequencies. 

In this work, we demonstrate coherent multi-spin exchange coupling in a GaAs quadruple quantum dot. We show that the nonlinear and non-local dependence of exchange couplings on confinement gate voltages results in large part from electronic wavefunction shifts during exchange pulses. We model our data using the Heitler-London expression for exchange coupling between two spins~\cite{DeSousa2001}, assuming that the barrier-gate pulses used to induce exchange coupling primarily shift the locations of the electrons. The model parameters we use change slightly depending on the number of spins involved, suggesting that additional effects beyond wavefunction shifts, including perhaps the quantum-dot potential depths and widths, are also important. 

The parameters we extract by fitting our data to the Heitler-London model agree well with electrostatic simulations of the confinement potential of our device. We also show that a simpler, exponential model also fits our data well and can be used to predict gate voltages for independent control of exchange couplings. We demonstrate two-, three-, and four-spin exchange coupling in our four-dot device. These results are applicable to Si qubits, which feature reduced hyperfine coupling and longer electron spin coherence compared to GaAs spin qubits. Our results are also applicable to longer arrays of spin qubits, an encouraging prospect for quantum information processing and the exploration of Heisenberg spin chain physics.
\begin{figure}
	\includegraphics{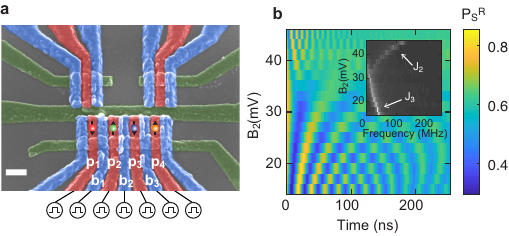}
	\caption{\label{Fig1} (a) Scanning electron micrograph of the quadruple quantum dot device. The scale bar is 200 nm. (b) Measured exchange oscillations with virtual barrier gate $B_3=30$ mV. Inset: Absolute value of the fast Fourier transform of data shown in (b). As $B_2$ increases, $J_{3}$ decreases.}
\end{figure}

\section{Device}
We use a quadruple quantum dot in a GaAs/AlGaAs heterostructure with overlapping-gates [Fig.~\ref{Fig1} (a)]~\cite{Angus2007,Zajac2015OverlapGates,Zajac2016Scalable}.  Two extra quantum dots placed above the main-dot array serve as charge sensors, and are configured for rf-reflectometry~\cite{Reilly2007FastRF,Barthel2010rfReadout}. We tune the confinement potential  using ``virtual gates''  ~\cite{Baart2016,Hensgens2017,Mills2019VirtualGates,Volk2019VirtualGates,Mills2019Computer} such that each dot contains only one electron. We define virtual plunger gate voltages $P_1$, $P_2$, $P_3$, and $P_4$ as linear combinations of the physical plunger gate voltages ($p_1$, $p_2$, $p_3$, $p_4$) such that changes to $P_i$ are proportional to changes in the electrochemical potential of dot $i$.  We also define virtual barrier-gate voltages $B_1$, $B_2$, and $B_3$ as the voltage applied to the corresponding physical barrier ($b_1$, $b_2$, and $b_3$) together with a linear combination of physical plunger voltages chosen such that the chemical potentials of the dots are unchanged by the barrier pulse. In particular, barrier gate pulses therefore involve voltages applied to physical barrier gates in addition to compensation pulses applied to all four physical plunger gates.  The virtual gates are related to the actual voltages via the measured capacitance matrix $\mathbf{A}$ through $\mathbf{G}=\mathbf{A}\cdot \mathbf{g}$, where $\mathbf{G}=[P_1, P_2, P_3, P_4, B_1, B_2, B_3]^T$ and $\mathbf{g}=[p_1,p_2, p_3, p_4, b_1, b_2, b_3]^T$. In the following, we will use the term ``virtual gate voltages" to mean pulses away from the idling tuning of the device, which is the symmetric operating point~\cite{Reed2016,Martins2016SymmetricExchange} of the four-dot array with one electron in each dot. 

For initialization and readout, we configure the four-spin array into two pairs. We refer to spins 1 and 2 as the left pair, and spins 3 and 4 as the right pair. We initialize the array in the product state $\ket{\D\U\D\U}$ via adiabatic separation of singlets in the hyperfine gradient~\cite{Petta2005STQubit,Kandel2019SpinTransfer}. Here the arrows indicate the spin states of all four spins. Alternatively, we can prepare a polarized triplet state $\ket{\U\U}$ in either pair by exchanging electrons with the reservoirs while each dot contains a single electron~\cite{Orona2018TpLoad}. We measure both pairs via Pauli spin blockade~\cite{Petta2005STQubit,Barthel2009RapidSingleShot,Kandel2019SpinTransfer} together with a shelving mechanism~\cite{Studenikin2012Readout21} to enhance the readout fidelity.

The spin-state Hamiltonian of the quadruple dot is
\begin{equation} \label{ham}
H = \frac{h}{4}\sum_{i=1}^{3}J_{i}(\boldsymbol{\sigma}_i\cdot\boldsymbol{\sigma}_{i+1}) + \frac{h}{2}\sum_{i=1}^{4}B^z_i \sigma^z_i.
\end{equation}
Here $J_{i}$ is the exchange coupling strength (with units of frequency) between dots $i$ and $i+1$, $\boldsymbol{\sigma}_i = [\sx_i, \sy_i, \sz_i]$ is the Pauli vector describing the components of spin $i$, and $h$ is Planck's constant. $B^z_i$ is the $z$-component magnetic field experienced by each spin, and it includes both a large 0.5 T external magnetic field and the smaller hyperfine field. The quantization axis ($z$-direction) is defined by the external magnetic field direction. The $x$- and $y$-components of the hyperfine field are neglected in this Hamiltonian since their sizes are negligible compared to the external magnetic field. $B^z_i$ also has units of frequency. 

\section{The effect of position shifts on exchange coupling}
A single non-zero exchange coupling $J_i$ is easily tuned by adjusting the voltage on the relevant barrier $B_i$~\cite{Reed2016,Martins2016SymmetricExchange}. When we extend the interaction to more than two spins by pulsing another virtual barrier gate $B_j$, however, the original exchange coupling $J_i$ is strongly affected. For example, a large pulse to $B_3$ nominally induces a nonzero $J_3$. But adding an additional pulse to $B_2$ during the evolution rapidly and nonlinearly reduces $J_3$ before eventually turning on a $J_2$  [Figs.~\ref{Fig1}(b)]. In fact, $J_3$ reduces to nearly zero before $J_2$ turns on. 

\begin{figure}
	\includegraphics{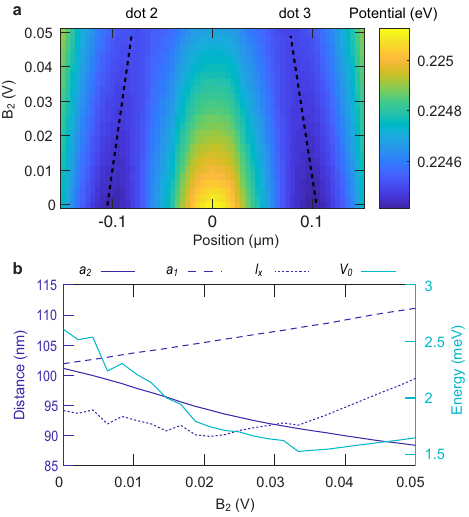}
	\caption{\label{Fig15} Electrostatic simulations. (a) Linecuts of the simulated potential associated with dots 2 and 3 vs. the barrier voltage pulse $B_2$. The left dip is the potential of dot 2, and the right dip is the potential of dot 3. The dots move closer together as $B_2$ increases. The dashed lines are guides to the eye. (b) Fitted parameters of the simulated double-dot potential vs. $B_2$. Based on the simulated potential of dot 1 (not shown), we also find that dots 1 and 2 move farther apart during the sample pulse.}
\end{figure}

We now show that this effect primarily results from lateral shifts of the quantum dots during a barrier-gate pulse. We have self-consistently calculated the electron density and potential of our device in COMSOL using the Thomas-Fermi approximation~\cite{Frees2019}. Our simulation replicates the behavior of the physical device with high fidelity. For example, the predicted incremental voltage on $p_4$ required to transition from one to two electrons in dot 4 is about 12.5 mV. This corresponds quite closely to the behavior of the device. The same transition in the physical device requires a change of 12 mV on $p_4$. Figure~\ref{Fig15}(a) shows the potential associated with dots 2 and 3 tuned to single-occupancy as a function of the barrier voltage $B_2$.  Our simulations include compensation pulses on the plunger gates to match our use of virtual gates in the actual experiment.
The dots clearly move toward each other as $B_2$ increases.

To obtain more detailed information about this process, we fit our two-dimensional simulated potentials to an equation of the form~\cite{DeSousa2001}:
\begin{eqnarray} \label{eq:pot}
V(x,y)&=&-V_0 \left[ \exp\left(\frac{-(x-a)^2}{l_x^2}\right)+\exp\left( \frac{-(x+a)^2}{l_x^2}\right) \right] \nonumber \\ 
& \times&\exp\left( \frac{y^2}{l_y^2}\right).
\end{eqnarray} 
Here $x$ and $y$ are coordinates in the plane of the two-dimensional electron gas, and $V_0$, $a$, $l_x$, $l_y$ characterize the potential wells of the dots. Double Gaussians of this type are commonly used to model double dots, but usually a separate barrier term is included, as in Ref. ~\cite{DeSousa2001}. Our simulated potential is shallow enough that a separate barrier term is not required to reproduce the potential we simulate. By fitting our simulated potentials to Eq. 2, we extract how the parameters $a$, $l_x$, and $V_0$ vary with $B_2$ [Fig.~\ref{Fig15}(b)]. The distances between dots 2 and 3, ($2a_2$) and dots 1 and 2 ($2a_1$) change approximately linearly during the barrier pulse, but in opposite directions, because dot 2 moves closer to dot 3 but farther from dot 1. Based on the simulations, we calculate that $a_2$ changes by about -0.3 $\mu$mV$^{-1}$, and $a_1$ changes by about 0.19 $\mu$mV$^{-1}$. Other parameters of the confinement potential change as well during the barrier pulse.

Reference~\cite{DeSousa2001} computes the exchange coupling between two quantum dots in a potential of the form Eq.~\ref{eq:pot} in the Heitler-London (HL) framework. At zero magnetic field, the result is 
\begin{eqnarray}
J_{HL}(V_0,a)&=&\frac{2 S^2}{1-S^4} \Bigg\{ \frac{\hbar^2 a^2 }{m l_0^4} -\frac{2 V_0 l_x l_y}{\sqrt{(l_x^2+l_0^2)(l_y^2+l_0^2)}} \nonumber \\
&\times& \left[ \exp \left( \frac{-(2a)^2}{l_x^2+l_0^2}\right)- 2 \exp \left( \frac{-a^2}{l_x^2 +l_0^2}\right)\right] \nonumber \\
&-&  \sqrt{\frac{\pi}{2}}\frac{e^2}{4 \pi \epsilon \epsilon_0 l_0}\left[1-S I_0\left(\frac{a^2}{l_0^2}\right)\right]\Bigg\}. \label{eq:HL}
\end{eqnarray} 
Here $S=\exp(-a^2/l_0^2)$, $l_0=\sqrt{\hbar/ m \omega_0}$, with $\omega_0=\sqrt{V_0/m l_x^2}$. $m$ is the electron effective mass, $\epsilon_0$ is the permittivity of free space, $\epsilon$ is the dielectric constant of the material, and $I_0$ is the zeroth order modified Bessel function. In writing this equation, we have assumed that the minima of the double-dot potential occur at $x=\pm a$. We have also ignored the magnetic-field-dependent terms, because for the magnetic field used here (0.5T), the effective magnetic confinement is still weaker than the electrostatic confinement. 

\begin{figure}
	\includegraphics{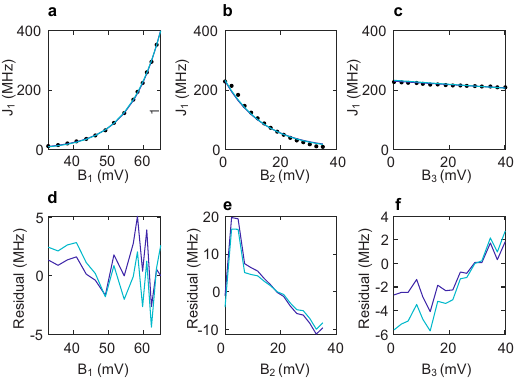}
	\caption{\label{Fig25} $J_1$ vs. $B_j$. (a) $J_1$ vs. $B_1$. (b) $J_1$ vs $B_2$, for $B_1 = 60$ mV. (c) $J_1$ vs. $B_3$, for $B_1 = 60$ mV. The black data points in each panel are obtained from the fast Fourier transform of a dataset similar to Fig.~\ref{Fig1}(b). In (a)-(c), the dark blue line is the fit to the exponential model, and the light blue line is the fit to the HL model. Panels (d)-(f)  show the difference between the fits and the data for the two models. }
\end{figure}

To determine if lateral position shifts can indeed explain our data, we experimentally measure how $B_j$ affects $J_1$ in our device (Fig.~\ref{Fig25}). We then fit our data to Eq.~\ref{eq:HL}. To parameterize the effect of the $B_j$, we allow for  $a_1=\alpha_{1}B_1+\alpha_{2}B_2+\alpha_{3}B_3$, where the $\alpha_{j}$ are fit parameters. We also fit for $V_0$, but we constrain $l_x=l_y=100$ nm, which is approximately the value we obtain from our simulations. The fitted values are $V_0=11.4$ meV, $\alpha_{1} = -0.40\mu$mV$^{-1}$, $\alpha_2=0.25\mu$mV$^{-1}$, and $\alpha_3=0.01\mu$mV$^{-1}$. The simulated and fitted values of $V_0$ are on the same order of magnitude. This level of agreement is reasonable, considering that the simulation calculates the semi-classical self-consistent potential associated with the electron density in the quantum dots. Moreover, the potential of each dot is not exactly Gaussian, making it challenging to extract an exact potential depth.   Our simulated values of the quantum-dot position shifts (-0.3 $\mu$mV$^{-1}$ and 0.19 $\mu$mV$^{-1}$) described above agree well with the fitted values of $\alpha_1$ and $\alpha_2$. This agreement supports our hypothesis that  lateral position shifts cause the observed trends in our data.

\section{Modeling the dependence of exchange coupling on all barrier gates}
The nonlinear and nonlocal dependence of exchange couplings on the barrier gate voltages, which results from position shifts of the quantum dots, poses a challenge to implementing simultaneous exchange coupling between all dots in an extended array. Previous work has investigated how to adjust multiple interdot tunnel couplings iteratively~\cite{Hensgens2017,Mills2019Computer}. Here, we discuss two different models, which allow us to determine the virtual gate voltages given a set of target exchange couplings. In contrast to previous iterative approaches, our approach generates a predictive model. As discussed further below, predictive tuning of exchange couplings in extended spin chains is especially helpful when multiple exchange couplings are present, because the observed spin oscillation frequencies do not correspond with the bare two-spin Heisenberg couplings. We also use our model to control coherent exchange coupling instead of incoherent electron tunneling. Finally, our approach has the advantage that it enables calibrating exchange couplings at the symmetric operating point, where tunnel couplings cannot easily be determined.

Our general approach is to measure how all of the $J_i$ depend on the $B_j$ and then to fit the parameters of a nonlinear model to the data. Using these fit parameters, we create a model which allows us to generate a set of virtual gate voltages $\mathbf{G}(\mathbf{j})$, for a set of target exchange coupling values $\mathbf{j}=[j_1, j_2, j_3]$, where the $j_i$ are the desired exchange coupling values. We compute the actual gate voltages using the transformation described above. We validate this model by inducing exchange coupling between two, three, and four spins and compare our observations with simulations, and we find good agreement.

To calibrate the models, we begin by inducing one strong exchange coupling $J_i \gg J_{j\neq i}$ and measuring the effect of the  $B_j$ on that exchange coupling. For example, to measure how $J_1$ depends on $B_j$, we initialize the array as discussed above with $B_j=0$. Then we pulse $B_1$ from 25 to 65 mV, and we record exchange oscillations at each pulse height. Setting $B_1 = 60$ mV, which yields a large but still-measurable $J_1 \sim 200$ MHz, and $B_3 = 0$ mV, we then pulse $B_2$ from 0 to 35 mV, and we record exchange oscillations. Setting $B_1 = 60$ mV and $B_2 = 0$ mV, we sweep $B_3$ from 0 to 40 mV and again record exchange oscillations. The pulses on $B_2$ are not sufficient to induce substantial $J_2$ due to the large pulse height on $B_1$. The pulses on $B_3$ do induce substantial $J_3$, but $J_1$ is not affected by the next-nearest-neighbor exchange coupling. We extract the oscillation frequencies through a fast Fourier transform of the data (Fig.~\ref{Fig25}). We repeat this process for the other $J_i$ (see Supplementary Material). 

The resulting $J_i$ vs $B_j$ data may be fitted to a set of equations related to Eq.~\ref{eq:HL}, of the form $J_i=J_{HL}(V_0^i,a_i)$, where
\begin{eqnarray}
a_1&=&a+\alpha_{11}B_1+\alpha_{12}B_2+\alpha_{13}B_3 \label{eq:hlmodel1} \\
a_2&=&a+\alpha_{21}B_1+\alpha_{22}B_2+\alpha_{23}B_3 \label{eq:hlmodel2}\\
a_3&=&a+\alpha_{31}B_1+\alpha_{32}B_2+\alpha_{33}B_3, \label{eq:hlmodel3}
\end{eqnarray}
where the $V_0^i$ and the $\alpha_{ij}$ are fit parameters. As discussed above, we constrain $l_x$ and $l_y$ to be the values found from simulations. These equations model our data quite well (Fig.~\ref{Fig25}), and the parameters we extract from the fits agree reasonably with our simulations. Values of $V_0^i$ range from 6.4 to 11.4 meV, and values of $\alpha_{ii}$ range from -0.40 to -0.43$\mu$mV$^{-1}$. Once the model is calibrated and the parameters found, we choose target exchange coupling values $\mathbf{j}$. We then numerically solve the set of equations $j_i=J(V_0^i,a_i)$ for the interdot separations $a_i$, and then we invert Eqs.~\ref{eq:hlmodel1}-\ref{eq:hlmodel3} to find the desired barrier gate voltages.

\begin{figure*}
	\includegraphics{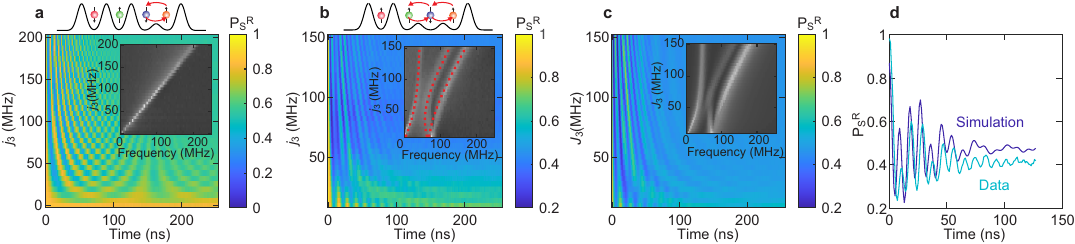}
	\caption{\label{Fig3} Two- and three-spin exchange coupling. (a)Two-spin exchange oscillations obtained by linearly sweeping $j_3$ from 0 to 200 MHz. Inset: FFT of the data. (b) Three-spin exchange oscillations obtained by linearly sweeping $j_3$ from 10 to 150 MHz and fixing $j_2$ at 70 MHz. Inset: FFT of the data. Theoretical predictions of the exchange oscillation frequencies are overlaid in red. (c) Simulated three-spin exchange oscillations corresponding to the data in (b). Inset: simulated FFT. (d) Linecuts from panels (b) and (c) at $j_2=j_3=70$ MHz.}
\end{figure*}

While this model (the ``HL model") originates from a microscopic theory, the exchange coupling is a highly non-linear function of the potential parameters, and some a-priori knowledge of the quantum-dot confinement potential is desirable. Using the HL model in practice also involves numerically solving non-linear equations, which can be susceptible to errors. An alternative, more robust, model for the dependence of the $J_i$ on the $B_j$ is motivated by the realization that the part of the expression for $J_i$ in Eq.~\ref{eq:HL} that is most sensitive to the inter-dot separation is the factor $S_i^2 \propto \exp{(-2a_i^2/l_0^2)}$. Setting $a_i = a + \eta_i$, where $\eta_i \ll a \approx 100$ nm, and $\eta_i \ll l_0 \approx 32$ nm we have 
\begin{eqnarray} \label{eq:exp}
J_i \sim \exp{(-2a_i^2/l_0^2)}\approx \exp{(-2a^2/l_0^2)}\exp{(-4a\eta_i/l_0^2)}. 
\end{eqnarray}
If $\eta_i=\alpha_{i1}B_1+ \alpha_{i2}B_2+\alpha_{i3}B_3$ as before, we expect the $J_i$ to depend approximately exponentially on the $B_j$. Thus, we introduce the following ``exponential model" for our data:
\begin{eqnarray}
\label{model1} J_{1} &= \beta_1\exp(\delta_{11}B_1+\delta_{12} B_2+\delta_{13}B_3) + \gamma_1 \label{eq:expmodel1} \\ 
\label{model2} J_{2} &= \beta_2\exp(\delta_{21}B_1+\delta_{22}B_2+\delta_{23}B_3) + \gamma_2   \label{eq:expmodel2} \\ 
\label{model3} J_{3} &= \beta_3\exp(\delta_{31}B_1+\delta_{32}B_2+\delta_{33}B_3) + \gamma_3 . \label{eq:expmodel3}
\end{eqnarray}
Here $\delta_{ij}$, $\beta_{i}$, $\gamma_i$ are fit parameters. Empirically, the fit parameters $\gamma_i$ are required for two reasons. As discussed above, the exchange couplings are not pure exponential functions of the barrier gates. Second, the hyperfine gradient can increase the measured oscillation frequency above the bare exchange frequency. Including $\gamma_i$ in the fit allows us to accommodate these deviations from pure exponential behavior.

The exponential model also matches our data quite well (Fig.~\ref{Fig25}). Typical values of $\beta_i$ and $\gamma_i$ are on the order of 10 MHz, and values of $\delta_{ii}$ range from 93 to 114 V$^{-1}$. From Eq.~\ref{eq:exp}, we expect that $\delta_{ij}\approx -4a\alpha_{ij}/l_0^2$. Using $a=100$ nm, $l_0=32$nm, and taking a typical value of $\alpha_{ii}=-0.4 \mu$mV$^{-1}$, we expect $\delta_{ii}\approx 156 $V$^{-1}$, which agrees reasonably well with our fitted values. We have also conducted measurements to confirm that the  $\delta_{ij}$ do not depend significantly on the barrier gate voltages, supporting the form of the exponential model (see Supplementary Material).

With the model parameters in hand, we choose a set of target exchange coupling values $\mathbf{j}$. Setting $\mathbf{J}=\mathbf{j}$, we invert Eqs. ~\ref{eq:expmodel1}-\ref{eq:expmodel3} to find the required virtual barrier gate voltages:
\begin{equation} \label{}
\begin{bmatrix}
B_1 \\
B_2 \\
B_3 \\
\end{bmatrix}
= 
\begin{bmatrix}
\delta_{11} & \delta_{12} & \delta_{13} \\
\delta_{21} & \delta_{22} & \delta_{23} \\
\delta_{31} & \delta_{32} & \delta_{33} \\
\end{bmatrix}
^{-1}
\begin{bmatrix}
\log\big((j_1-\gamma_1)/\beta_1\big) \\
\log\big((j_2-\gamma_2)/\beta_2\big) \\
\log\big((j_3-\gamma_3)/\beta_3\big) \\
\end{bmatrix} \,.
\end{equation}

For both the exponential and HL models, we require that the virtual plunger gates remain fixed at the symmetric operating point, and then we transform the virtual gate voltages to physical gate voltages using the capacitance matrix through $\mathbf{g}=\mathbf{A}^{-1}\cdot \mathbf{G}$, as discussed above.

\section{model validation}

\begin{figure*}
	\includegraphics{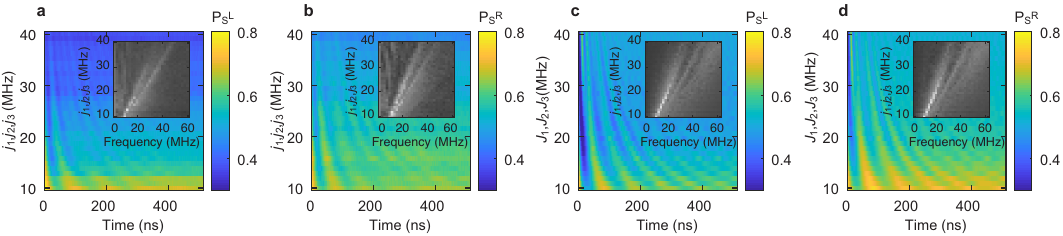}
	\caption{\label{Fig4} Four-spin exchange oscillations. (a) Experimental data measured on the left side. Inset: FFT of the data. (b). Experimental data measured on the right side. Inset: FFT of the data. (c) Simulated four-qubit exchange oscillation data on the left side. Inset: FFT of the simulation. (d) Simulated four-qubit exchange oscillation data on the right side. Inset: FFT of the simulation.}
\end{figure*}

In practice we prefer to use the exponential model because it features a robust inversion process and fits our data as well as the HL model (Fig.~\ref{Fig25}). (See the Supplementary Material for a comparison of the gate voltages generated by these two models.) We first validate our approach by sweeping $j_3$, the target exchange coupling between spins 3 and 4, linearly from 0 to 200 MHz [Fig.~\ref{Fig3}(a)]. The observed oscillation frequency matches our expectation. We also perform the same test on qubits 1 and 2, and qubits 2 and 3, and observe good agreement between the measured and target frequencies. 

In our device, we estimate that the residual exchange coupling at the idling point with no barrier-gate pulse is not zero but rather a few megahertz for each pair. In generating the data of Fig.~\ref{Fig3}(a), which features a swept $j_3$, we did not simultaneously require that $j_1=j_2=0$. If we had required $j_1=j_2=0$, our model would generate large negative values of $B_1$ and $B_2$. Instead, for this data set, we fixed $B_1=B_2=0$, since the weak residual exchange coupling does not significantly affect these data.

Next we induce three-spin exchange oscillations between spins 2, 3, and 4. The array is initialized in the $\ket{\U\U\D\U}$ state. Preparing $\ket{\U\U}$ on the left side ensures that spins 2, 3, and 4 remain in the $s_z=+\frac{1}{2}$ subspace, regardless of the sign of the local magnetic field gradient. We fix $B_1=0$, $j_2=70$ MHz, and we sweep $j_3$ linearly from 10 to 150 MHz, and we measure oscillations on the right pair. We compare our data to simulated predictions (see Supplementary Material for details on the simulation). The observed frequencies closely match our expectation, confirming that we can correctly set the target frequencies [Figs.~\ref{Fig3}(b)-(c)]. Note the presence of three distinct frequencies in the measured spectrum of Fig.~\ref{Fig3}(b). These frequencies are not the bare Heisenberg couplings. Instead, they result from the different energy splittings between the singlet-like and triplet-like states of three spins~\cite{Laird2010ExchangeOnly}. The theoretically predicted~\cite{Laird2010ExchangeOnly} low-lying energy splittings of three exchange-coupled spins are shown in red in Fig.~\ref{Fig3}(b), and they overlap nicely with our measurements~\cite{Laird2010ExchangeOnly}. This theoretical prediction assumes zero magnetic gradient between the dots. The presence of a hyperfine gradient in our device causes the experimental data to slightly deviate from the theoretical predictions [Fig.~\ref{Fig3}(c)]. However, this deviation is relatively insignificant for exchange strengths above 10 MHz. 

The model parameters we extract from the fits alone do not suffice to accurately generate the target three-spin exchange frequencies for both the exponential and HL models. We empirically find that the voltages $\mathbf{G}(\mathbf{j})$ generate actual exchange frequencies $\mathbf{J}<\mathbf{j}$ when two or more of the $j_i$ are non-zero (see Supplementary Material). To overcome this challenge, we make minor adjustments to the fitted model parameters and compare the observed three-spin exchange frequencies with simulations. We repeat this process for a few iterations until the experimental data match closely the simulated predictions, usually within about 10 MHz. In the exponential model, we normally need to modify the values of $\beta_{i}$ and $\gamma_{i}$, while the values of $\delta_{ij}$ can remain mostly unchanged. For the HL model, we usually need to increase slightly the confinement strength $V_0$. See the Supplementary Material for a comparison of the fitted and adjusted parameters and gate voltages.

The need for a modified parameter set for three-spin exchange coupling may originate for the following reasons. First, we calibrate our model when only one of the $J_i$ is large. However, three-spin exchange coupling requires multiple large $J_i$, and this requires several large, simultaneous barrier-gate pulses. Second, our assumption that only linear position shifts induce exchange coupling likely breaks down at large gate voltages. Indeed, Fig.~\ref{Fig15}(b) shows that both the characteristic size and confinement energy change during barrier gate pulses. Third, generating three-spin exchange coupling involves simultaneous large voltage pulses on several plunger and barrier gates, and any errors in our measured capacitance matrix will cause errors in the exchange couplings.  

Despite the need for an additional set of parameters for three-spin exchange coupling, the data in Fig.~\ref{Fig3} show that our model can still be used in this regime. Indeed, this additional set of parameters also suffices to induce four-spin exchange coupling, because the exchange coupling has vanishing dependence on the next-nearest-neighbor barrier gates [Figs.~\ref{Fig25}(c)]. To calibrate the two sets of three-spin parameters needed for four-spin coupling, we first tune the model for exchange coupling between spins 2-3-4, which yields precise values of $\alpha_{22}$, $\alpha_{23}$, $\alpha_{32}$, and $\alpha_{33}$, together with $\beta_2$, $\beta_3$, $\gamma_2$, and $\gamma_3$. To calibrate the model for exchange coupling between spins 1-2-3, we leave these parameters fixed, and tune $\alpha_{11}$, $\alpha_{12}$, $\alpha_{21}$, $\beta_1$, and $\gamma_1$. With the parameters tuned in this way, we induce simultaneous exchange coupling between all four spins in the array by initializing the array in the $\ket{\D\U\D\U}$ state, and we sweep all three target exchange frequencies linearly from 10 to 40 MHz. The oscillations are measured on both the left and the right pairs [Figs.~\ref{Fig4}(a)-(b)]. The experimental data and the simulated predictions [Figs.~\ref{Fig4}(c)-(d)] match closely. The agreement between the experimental data and the simulation shows we have good control over all exchange couplings. The maximum simultaneous exchange coupling is limited to about 40 MHz, because larger barrier pulses disrupt the tuning of the device. 

\section{Discussion}

A significant source of error in our model is the fluctuating nuclear hyperfine field, which is of order 10 MHz in our device (see Supplementary Material). Empirically, our model can generate exchange couplings which are accurate to about 10 MHz, suggesting that the hyperfine fields are a significant source of error. Our model becomes increasingly inaccurate when exchange frequencies approach the hyperfine field values. 

Although different sets of parameters are needed for two- and three-spin exchange coupling, the range of validity of the model is sufficient to accurately predict voltages for a wide range of exchange coupling values in either of these cases. To improve on this model, additional effects beyond lateral position shifts probably need to be considered. As discussed above, it is likely that other parameters of the dots are changing or that the capacitance matrix may require further refinement. More sophisticated modeling of the electrostatic potential, using the actual computed potential, as opposed to a Gaussian approximation, and a more accurate calculation of exchange couplings beyond the two-spin HL framework ~\cite{Pan2020} will likely also help to improve the model.

In the future, it seems likely that fabricating devices with extremely narrow barrier gates may help to reduce the voltages needed for simultaneous exchange coupling between multiple electrons in quantum-dot spin chains, although large pulses may still be required to adjust the exchange coupling from very small to very large values. It is less clear that the challenge of position shifts can be avoided with narrow barrier gates. In any case, we expect that electrostatic simulations can be used effectively to guide the design of future quantum-dot spin chains. We also expect that the use of electrostatic simulations to guide quantum-dot spin-qubit tuning and operation will become a valuable tool as quantum-dot devices increase in complexity.

We expect that our technique can apply to different types of quantum dots. A similar exponential dependence of exchange has been observed in Si/SiGe quantum dots~\cite{Reed2016,Zajac2017CNot} and GaAs devices~\cite{Martins2016SymmetricExchange}. The form of Equation ~\ref{eq:HL} suggests that lateral wavefunction shifts may be a primary cause of this behavior. A similar exponential model has also been used to describe the behavior of tunnel couplings in a GaAs device with a non-overlapping, open gate architecture~\cite{Hsiao2020}. We also expect that this model could be applied to Si/SiO$_2$ devices, where voltage-controlled tunnel coupling between quantum dots has now been demonstrated~\cite{Eenik2019}. 

The possibility of creating quantum-dot Heisenberg spin chains opens up a wide array of new phenomena to explore. As we have shown in previous work, pulsed exchange couplings between spins in quantum-dot spin chains can enable new forms of quantum information transfer~\cite{Kandel2019SpinTransfer,Qiao2019}. When multiple exchange couplings can be controlled simultaneously, coherent information transfer via antiferromagnetic spin chains becomes a possibility~\cite{Bose2003SpinChain,Bose2007,Friesen2007SpinBus}. Various techniques related to adiabatic quantum information processing, including adiabatic state transfer~\cite{Greentree2004} and adiabatic state and gate teleportation~\cite{Bacon2009} seem within reach. Single-pulse three-qubit gates~\cite{Gullans2019Toffoli} and the creation of long-range entangled states also become possible. Beyond quantum computing applications, Heisenberg spin chains are essential elements of models for quantum magnetism, and they underlie important non-equilibrium many-body phenomena of great interest, such as effects related to many-body localization~\cite{Pal2010,Barnes2016MBL} and time crystals~\cite{Barnes2019TimeCrystal}. 

As an example of the utility of our approach, we return to the experiment described in Fig.~\ref{Fig3}, where we generate coherent exchange coupling between three spins. These data represent an experiment in which an initial state $\ket{\U \U \D \U}$ evolves under simultaneous exchange coupling between dots 2-3 and 3-4. When $J_2=J_3$, evolution for a total time $\frac{2}{9 J_2}$ yields the final state $\ket{\psi}=\frac{1}{\sqrt{3}} \ket{\U} \otimes \left( \ket{\U \U \D} + e^{i 2\pi/3} \ket{\U \D \U} +\ket{\D \U \U} \right)$. $\ket{\psi}$ is equivalent to a W state on qubits 2, 3, and 4, up to single-qubit rotations. For the case shown in Fig.~\ref{Fig3}(d), $j_2 =j_3 \approx$ 70 MHz. Thus, in a total evolution time of approximately 3.2 ns, we expect to have generated a three-qubit entangled state. Based on Monte Carlo simulations, which integrate the Schr\"odinger equation including initialization errors, charge noise, and hyperfine noise [Fig.~\ref{Fig3}(d)] (see Supplementary Material), we expect that we can prepare qubits 2, 3, and 4 in the state $\frac{1}{\sqrt{3}} \left( \ket{\U \U \D} + e^{i 2\pi/3} \ket{\U \D \U} +\ket{\D \U \U} \right)$ with a fidelity of about 0.60. The dominant source of infidelity is state preparation error associated with thermal population of excited states and the fluctuating hyperfine gradient. Neglecting preparation errors, we expect that the fidelity can exceed 0.9, largely because the total evolution time is much shorter than the relevant dephasing times in this system. The creation of extended W states in larger Heisenberg spin chains is discussed further in Ref.~\cite{Friesen2007SpinBus}.

In addition to generating a many-body entangled state, the case where $J_2=J_3$ can also generate a remote entangling operation, as we now discuss. An initial state $\ket{\U \U \U \D}$ evolves after a time $\frac{2}{3 J_2}$ to $\ket{\psi}=\frac{\sqrt{3}}{2}\ket{\U}\otimes\left(\ket{\D\U\U}+\frac{e^{i \pi/2}}{\sqrt{3}}\ket{\U \U \D}\right)$. $\ket{\psi}$ features entanglement only between qubits 2 and 4. Moreover, the initial state of qubit 3 is unchanged. In fact, it is easily verified that in the absence of magnetic gradients, this operation is equivalent to an $(S_{24})^{2/3}$ operation, where $S_{24}$ indicates a SWAP gate between qubits 2 and 4. Compare this to the common entangling operation $(S_{ij})^{1/2}$ between two qubits $i$ and $j$. Based on our simulations, we expect that with perfect state preparation, we could generate this three-qubit state with a fidelity of about $0.86$ in a total time of about 9.5 ns. The generation of remote exchange in extended exchange-coupled spin chains is further discussed in Ref.~\cite{Oh2011}.

\section{Conclusion}
In summary, we have demonstrated simultaneous coherent exchange coupling between two, three, and four spins in a quadruple quantum dot. We have also shown that lateral position shifts of the quantum dots during barrier pulses present a significant hurdle to implementing simultaneous exchange coupling between multiple electron spins. Using a phenomenological model based on a microscopic theory, we can predict the virtual gate voltages required to generate a set of target exchange couplings. Our method is also scalable and applicable to other types of quantum-dot spin qubits, such as silicon qubits, which offer the possibility of isotopic purification and substantially reduced nuclear hyperfine fields, leading to longer electron-spin coherence times.  This method enables us to generate a four-site Heinseberg spin chain, which is an exciting prospect for the exploration of the physics associated with interacting spin chains.

During the completion of this work, we became aware of a related work demonstrating independent control of tunnel couplings in a quantum dot array~\cite{Hsiao2020}.

\section{Acknowledgments}
This research was sponsored by the Defense Advanced Research Projects Agency under Grant No. D18AC00025, the Army Research Office under Grant Nos. W911NF16-1-0260 and W911NF-19-1-0167, and the National Science Foundation under Grant No. DMR-1941673. The views and conclusions contained in this document are those of the authors and should not be interpreted as representing the official policies, either expressed or implied, of the Army Research Office or the U.S. Government. The U.S. Government is authorized to reproduce and distribute reprints for Government purposes notwithstanding any copyright notation herein.

%

\end{document}


\renewcommand{\figurename}{Supplementary Figure}
\renewcommand{\thefigure}{\arabic{figure}}
\renewcommand{\tablename}{Supplemntary Table}

\title{Supplementary Material for\\Coherent multi-spin exchange coupling in a quantum-dot spin chain}

\author{Haifeng Qiao}
\thanks{These authors contributed equally.}

\author{Yadav P. Kandel}
\thanks{These authors contributed equally.}

\affiliation{Department of Physics and Astronomy, University of Rochester, Rochester, NY, 14627 USA}

\author{Kuangyin Deng}
\affiliation{Department of Physics, Virginia Tech, Blacksburg, Virginia 24061, USA}

\author{Saeed Fallahi}
\affiliation{Department of Physics and Astronomy, Purdue University, West Lafayette, IN, 47907 USA}
\affiliation{Birck Nanotechnology Center, Purdue University, West Lafayette, IN, 47907 USA}

\author{Geoffrey C. Gardner}
\affiliation{Birck Nanotechnology Center, Purdue University, West Lafayette, IN, 47907 USA}
\affiliation{School of Materials Engineering, Purdue University, West Lafayette, IN, 47907 USA}

\author{Michael J. Manfra}
\affiliation{Department of Physics and Astronomy, Purdue University, West Lafayette, IN, 47907 USA}
\affiliation{Birck Nanotechnology Center, Purdue University, West Lafayette, IN, 47907 USA}
\affiliation{School of Materials Engineering, Purdue University, West Lafayette, IN, 47907 USA}
\affiliation{School of Electrical and Computer Engineering, Purdue University, West Lafayette, IN, 47907 USA}

\author{Edwin Barnes}
\affiliation{Department of Physics, Virginia Tech, Blacksburg, Virginia 24061, USA}

\author{John M. Nichol}
\email{john.nichol@rochester.edu}

\affiliation{Department of Physics and Astronomy, University of Rochester, Rochester, NY, 14627 USA}


\pacs{}

\maketitle

\subsection{Device}
The quadruple quantum dot device is fabricated on a GaAs/AlGaAs heterostructure, with three layers of overlapping Al confinement gates. An additional grounded top gate covers the main device area for smoothing the potential anomalies caused by fabrication defects. The two-dimensional electron gas (2DEG) resides at the GaAs and AlGaAs interface, 91 nm below the semiconductor surface. The 2DEG density $n=1.5 \times 10^{11}$cm$^{-2}$ and mobility $\mu=2.5 \times 10^6$cm$^2/$Vs were measured at $T=4$K. The device is cooled in a dilution refrigerator with base temperature of approximately 10mK. An external magnetic field $B=0.5$T is applied parallel to the 2DEG, normal to the axis connecting the quantum dots.

\subsection{Measurement of Hyperfine Gradient}
\begin{figure*}
	\includegraphics{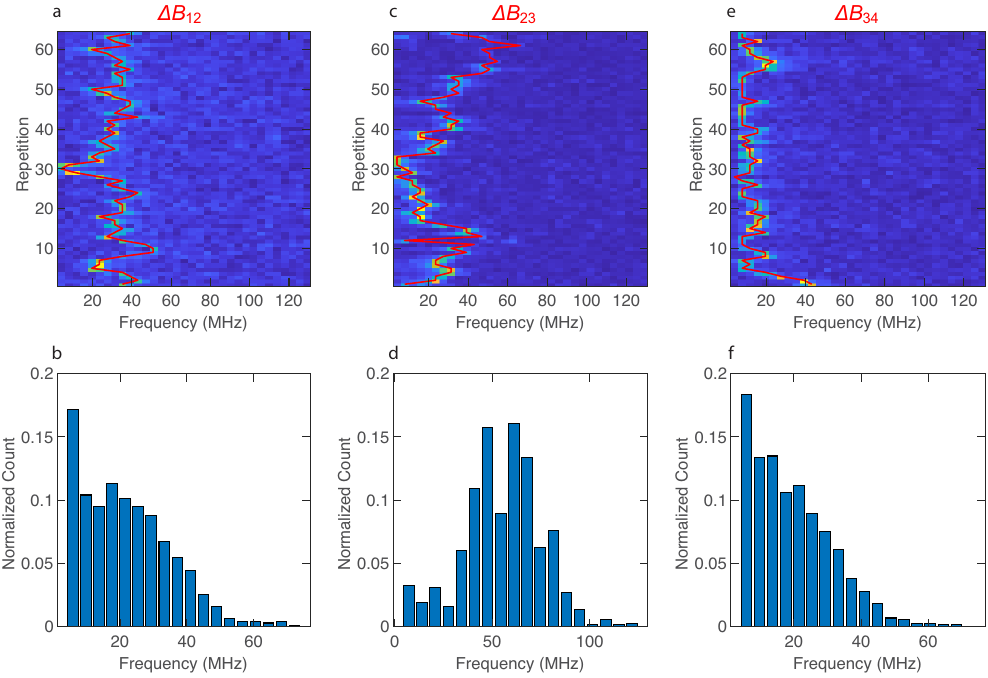}
	\caption{\label{SFig1} Measurements of hyperfine gradients. \tb{a} Absolute values of the FFT of $\Delta B_{12}$ oscillations, with extracted frequencies shown in red. Only 64 out of 2048 repetitions are shown. \tb{b} Extracted $\Delta B_{12}$ frequency distribution. \tb{c} Absolute values of the FFT of $\Delta B_{23}$ oscillations. \tb{d} $\Delta B_{23}$ frequency distribution. \tb{e} Absolute values of the FFT of $\Delta B_{34}$ oscillations. \tb{f} Extracted $\Delta B_{34}$ frequency distribution.}
\end{figure*}

We can measure the hyperfine gradients between neighboring quantum dots by inducing singlet-triplet oscillations and measuring the oscillation frequencies. First we prepare $\ket{S}=\frac{1}{\sqrt{2}}\left(\ket{\U\D}-\ket{\D\U}\right)$ states on both the left and the right sides in the (2,0,0,2) charge configuration, and diabatically separate them into the (1,1,1,1) configuration. The rapid charge separation allows both sides to retain their singlet states (rather than evolving to product states). The left side undergoes singlet-triplet oscillations with frequency $\Delta B_{12} = |B^{z}_{1} - B^{z}_{2}|$, and the right side oscillates with frequency $\Delta B_{34} = |B^{z}_{3} - B^{z}_{4}|$. $B^{z}_{i}$ is defined as in Eq. (1) in the main text. After both sides evolve for a variable amount of time around the gradients, we project them onto the $\{\ket{S},\ket{T}\}$ basis via diabatic charge transfer back to the (2,0,0,2) configuration. The measurements on the left side show singlet-triplet oscillations under $\Delta B_{12}$, and the measurements on the right side show oscillations under $\Delta B_{34}$. To measure the gradient $\Delta B_{23}$, we create an entangled state between qubits 2 and 3 via a $\sqrt{\text{SWAP}}$ gate~\cite{Kandel2019SpinTransfer}. We first initialize the array into the $\ket{\U\U\D\U}$ state, and then a $\sqrt{\text{SWAP}}$ gate between qubits 2 and 3 generates the entangled state $\frac{1}{\sqrt{2}}\left(\ket{S_{23}}+i\ket{T_{0,23}}\right)$, which undergoes singlet-triplet oscillations with frequency $\Delta B_{23}=|B^{z}_{2} - B^{z}_{3}|$, while qubits 1 and 4 remain unchanged. After some evolution time, another $\sqrt{\text{SWAP}}$ is applied between qubits 2 and 3 to preserve the coherence. The array is then adiabatically transferred back to the (2,0,0,2) configuration, and the oscillations are measured on the right side. Each data point is averaged for 128 iterations, and we repeat the measurement 2048 times. We extract the oscillation frequencies from the FFT of the data, and we create histograms of the measured frequencies [Supplementary Fig.~\ref{SFig1}]. From the histograms we can obtain the hyperfine gradient fluctuations. In our device, the mean values of the gradient fluctuations appear to slowly drift over time, but the standard deviations remain relatively constant. We incorporate the experimentally obtained standard deviations to the simulation, but adjust the mean values in the simulation to better match the experimental data.

\subsection{Simulation}
We generate all simulated data by numerically integrating the Schr\"{o}dinger equation and calculating the singlet return probability for both the left side (qubits 1 and 2) and the right side (qubits 3 and 4). The Hamiltonian is given in Eq (1) in the main text. Our simulation includes errors from state preparation, relaxation during readout, readout fidelity, charge noise, and hyperfine field fluctuations. To simulate the errors in singlet load and adiabatic charge separation, we create the two-electron state
\begin{equation}
\ket{\tilde{S}} = s_{1}\ket{S} + s_{2}\ket{T_0} + s_{3}\ket{T_{+}} + s_{4}\ket{T_{-}} \,,
\end{equation}
where $|s_{1}|^2 = f_{s}$ is the singlet load fidelity, and $|s_{2}|^2 = |s_{3}|^2 = |s_{4}|^2 = \frac{1}{3}(1-f_{s})$. Here $\ket{T_0}=\frac{1}{\sqrt{2}}\left(\ket{\U \D}+\ket{\D \U} \right)$, $\ket{T_+}=\ket{\U \U}$, and $\ket{T_-}=\ket{\D \D}$. The error associated with the process of charge separation is neglected, since the dominating error source is the singlet loading error. In the case of the $T_{+}$ load, we simulate the state to be 
\begin{equation}
\ket{\tilde{T}} = t_{1}\ket{S} + t_{2}\ket{T_{0}} + t_{3}\ket{T_{+}} + t_{4}\ket{T_{-}} \,,
\end{equation} 
where the coefficients are calculated following the process described in Ref.~\cite{Orona2018TpLoad}. In both cases, random phases are assigned to the coefficients for each realization of the simulation. To simulate the errors from relaxation and readout fidelity, we calculate the final singlet return probability for each side to be
\begin{equation}
\tilde{P_{S}} = (1-g-2r)P_{S}+g+r \,,
\end{equation}
where $P_{S}$ is the true singlet return probability after evolution, $r = 1-f_{m}$ is the probability of misidentifying singlet as triplet (or vice versa) as a result of noise, and $g = 1-\exp(-t_{m}/T_{1})$ is the probability of the triplet relaxing to the singlet during measurements. Here $f_{m}$ is the measurement fidelity, $t_{m}$ is  the measurement time, and $T_{1}$ is the relaxation time. To account for charge noise and hyperfine field fluctuations, we vary the values of $J_{i}$ and $B^z_{i}$ in the Hamiltonian between simulation runs. The values in each run are sampled according to a Gaussian distribution. Each set of simulated data is averaged over 4096 realizations of charge noise, hyperfine field fluctuations, and state preparation errors.

To extract a fidelity for the three-qubit entangled states as discussed in the main text, we integrated the Schr\"odinger equation as discussed above. In this case, we traced out qubit 1 to obtain a reduced density matrix $\rho_{234}$ for qubits 2,3, and 4. We computed the fidelity as $\left(\textnormal{Tr} \sqrt{\sqrt{\rho} \rho_{234} \sqrt{\rho}} \right)^2$, where $\rho$ is the target density matrix defined by the states discussed in the main text. 

\subsection{Exchange Coupling Model}
\begin{figure*}
	\includegraphics{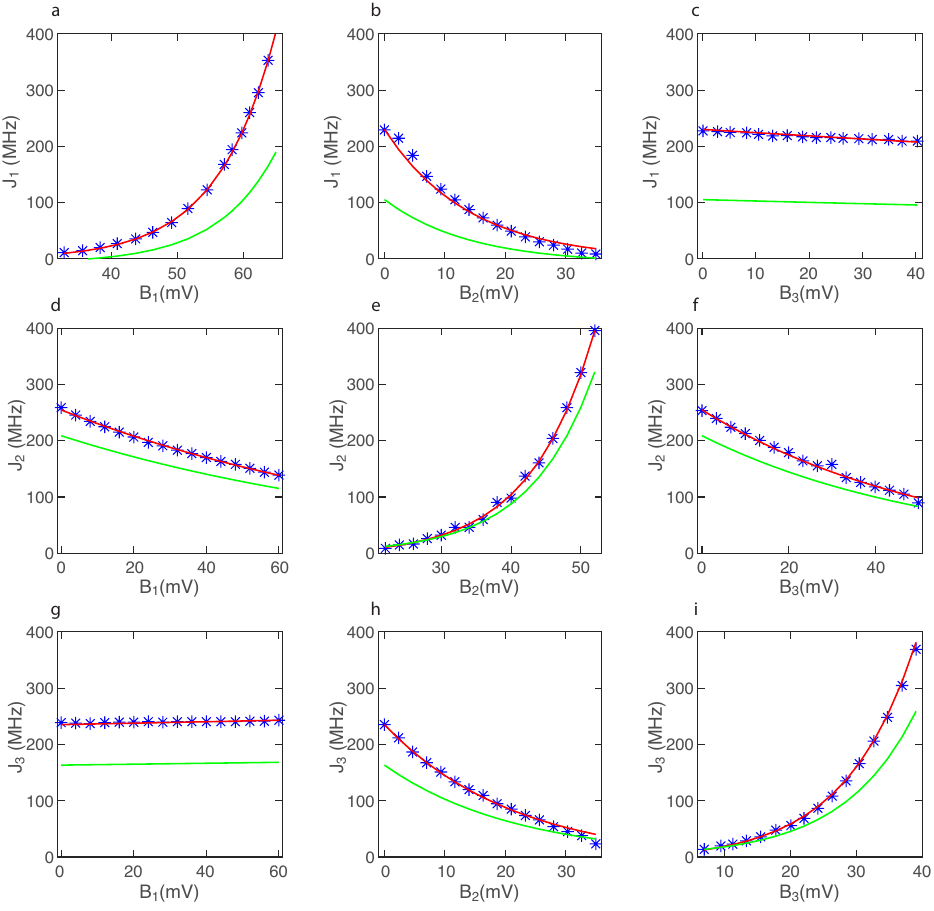}
	\caption{\label{SFig2} Full calibration data, initial, and modified fits for the exponential model. Blue data points are experimentally measured two-qubit exchange oscillation frequencies, red lines are initial exponential fit, and green lines are predicted frequencies using the model with modified parameters.}
\end{figure*}

We use the exponential model to describe the dependence of the exchange couplings on all barrier gate voltages (Supplementary Fig.~\ref{SFig2}). The parameters are obtained by fitting Eqs. (8)-(10) in the main text to exchange oscillation frequencies extracted from the calibration data. The model with these initial parameters accurately predicts the oscillation frequencies in two-qubit exchange interactions, but fails to do so for three- and four-qubit interactions. To obtain a set of parameters that is suitable for multi-qubit exchange oscillations, we modify the initial parameters by comparing the three-qubit exchange-oscillation data with the simulation. Eventually we have two sets of parameters: one for two-qubit exchange oscillations, and the other one for multi-qubit exchange oscillations. A comparison between the two sets of parameters are shown in Supplementary Table~\ref{tab1} and Supplementary Fig.~\ref{SFig2}. 

We show the robustness of the exponential model parameters in Supplementary Fig.~\ref{SFig3}. We induce the exchange coupling $J_{3}$ by increasing $B_3$, and at the same time we fix $B_2$ to a non-zero value (without inducing $J_{2}$). We repeat the experiment while varying $B_2$, and fit $J_{3}$ to an exponential function of $B_3$ and obtain $\delta_{33}$ for each value of $B_2$. Finally we monitor how $\delta_{33}$ changes with $B_2$. We also run the same test for $\delta_{32}$ and monitor how it changes with $B_3$ (Supplementary Fig.~\ref{SFig3}). Both $\delta_{32}$ and $\delta_{33}$ show minimal dependence on the nearest-neighboring barrier gate voltages, which indicates the functional form for our model is justified.

Table~\ref{tab1} shows typcal fitted and modified parameters for the exponential model.

\begin{figure*}
	\includegraphics{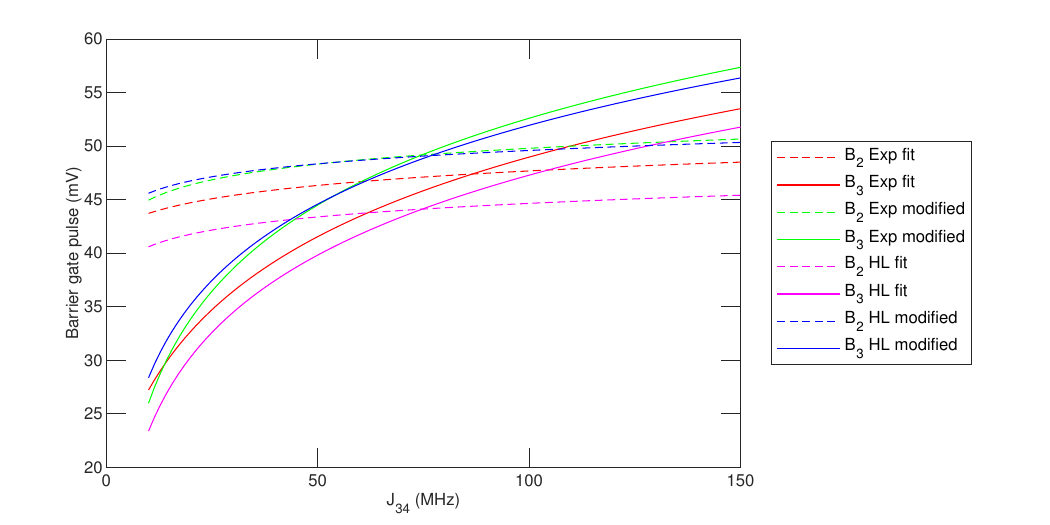}
	\caption{\label{SFig5} Comparison of gate voltages predicted by initial and modified fits for both the exponential and HL models. The red solid and dashed lines are the gate voltages for $B_2$ and $B_3$ generated by the exponential model for the three-spin exchange-oscillation data presented in the main text. The gate voltages were calculated using the parameters we obtain after fitting the calibration data. However, as discussed in the main text, these gate voltages generate exchange couplings that are too small. Therefore, we modify the model parameters as discussed in the main text, and the actual gate voltages we used are shown in the green solid and dashed lines. The magenta and blue lines show the gate voltages predicted by the initial and modified parameter sets for the HL model for the same experiment. The modified parameters for both the HL and exponential (blue and green lines) models generate rather similar gate voltages.}
\end{figure*}

Supplementary Fig.~\ref{SFig5} shows all fits to the calibration data using the HL model. 

%

\newpage

\begin{table*}
\centering
\begin{tabular}{|c|c|c|c|}
\hline
parameters & $\beta$ (MHz) & $\delta \times 10^{-3}$ (V$^{-1}$) & $\gamma$ (MHz) \\ [6pt]
\hline
initial &
$
\begin{bmatrix}
0.244  \\ 
1.424 \\
10.17
\end{bmatrix}
$
&
$
\begin{bmatrix}
0.1142 & -0.0717 & -0.0026 \\ 
-0.0099 & 0.1085 & -0.0184 \\
0.0005 & -0.0468 & 0.0932
\end{bmatrix}
$
&
$
\begin{bmatrix}
-0.604 \\ 
-4.993 \\
-6.882
\end{bmatrix}
$ \\
\hline
modified & 
$
\begin{bmatrix}
0.125 \\ 
1.15 \\
7.5
\end{bmatrix}
$
&
$
\begin{bmatrix}
0.1135 & -0.0713 & -0.0022 \\ 
-0.0099 & 0.1084 & -0.0184 \\
0.0005 & -0.0462 & 0.0907
\end{bmatrix}
$
&
$
\begin{bmatrix}
-8 \\ 
-0.25 \\
-0.5
\end{bmatrix}
$ \\ 
\hline
\end{tabular}
\caption{Comparison of the initial and the modified parameters used for modeling.}
\label{tab1}
\end{table*}

\begin{figure*}
	\includegraphics{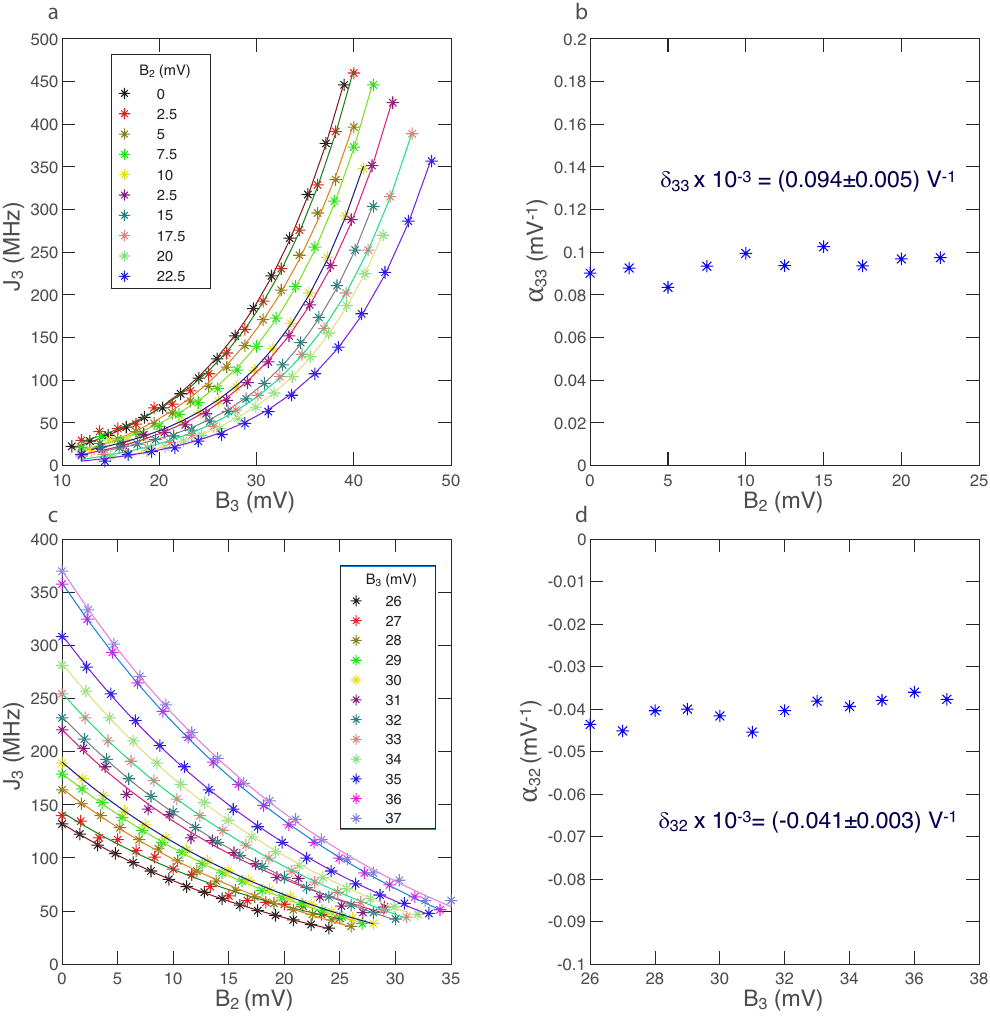}
	\caption{\label{SFig3} Robustness of exponential model parameters. \tb{a} Dependence of $J_{3}$ on the barrier $B_3$, with different values of $B_2$. Data points are experimentally obtained frequencies, and solid lines are exponential fits. \tb{b} Change in the fit parameter $\alpha_{33}$ with $B_2$. \tb{c} Dependence of $J_{3}$ on the barrier $B_2$, with different values of $B_3$. \tb{d} Change in the fit parameter $\alpha_{32}$ with $B_3$.}
\end{figure*}

\begin{figure*}
	\includegraphics{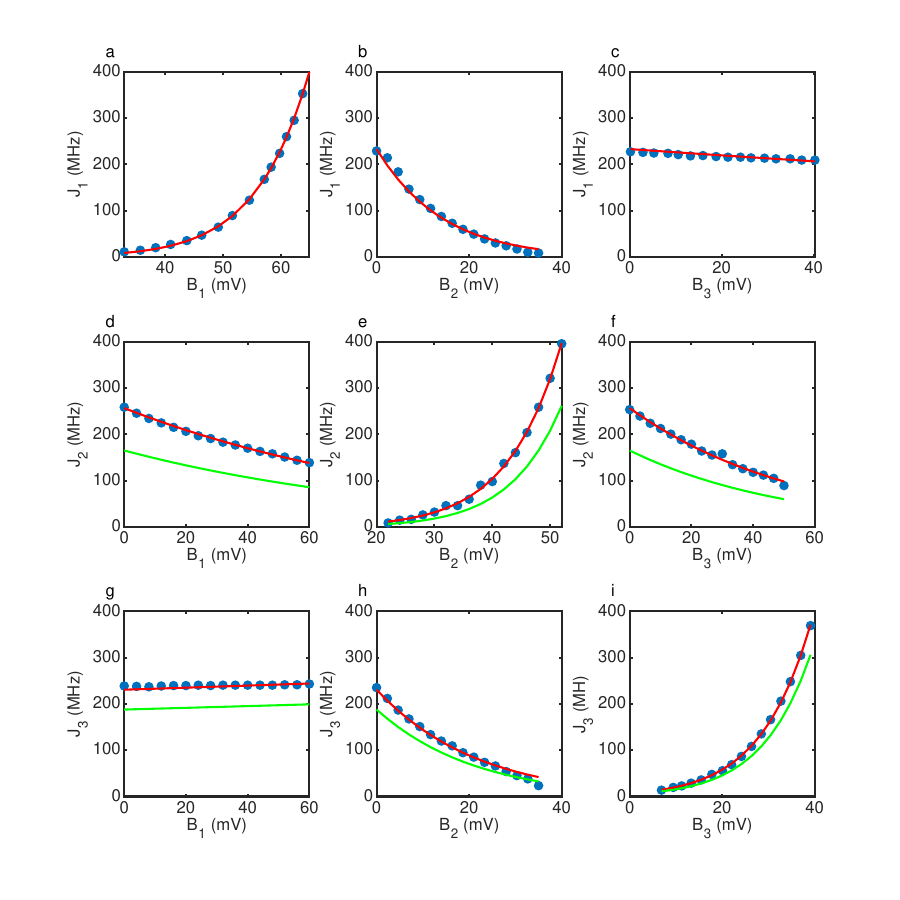}
	\caption{\label{SFig4} Full calibration data, initial, and modified fits for the HL model. Blue data points are experimentally measured two-qubit exchange oscillation frequencies, red lines are initial exponential fit, and green lines are predicted frequencies using the model with modified parameters. To generate the modified parameters, we adjusted the HL parameters to generate voltages to match the voltages used in the three-spin exchange-oscillation experiment. We only adjust parameters for $J_2$ and $J_3$. }
\end{figure*}

